\documentstyle[iopconf1,epsf]{article}
\begin{document}
\bibliographystyle{unsrt}
\title{The DRIFT Project: Searching for WIMPS with a Directional Detector}
\author{M. J. Lehner\dag\footnote{E-mail: m.lehner@sheffield.ac.uk},
K. Griest\ddag,
C. J. Martoff\S,
G. E. Masek\ddag,
T. Ohnuki$\|$,
D. Snowden-Ifft$\|$,
N. J. C. Spooner\dag}

\affil{\dag\ Department of Physics and Astronomy, University of Sheffield,
Sheffield~S3~7RH,~UK}

\affil{\ddag\ Department of Physics, University of California San Diego,
La~Jolla,~CA~92093-0319,~USA}

\affil{\S\ Department of Physics, Temple University,
Philadelphia,~PA~19122-6082,~USA}

\affil{$\|$\ Department of Physics, Occidental College,
Los~Angeles,~CA~90041,~USA}

\beginabstract
 A low pressure time projection chamber for the detection of WIMPs
is discussed.  Discrimination against Compton electron background in such a 
device should be very good, and directional information about the recoil 
atoms would be obtainable.  If 
a full 3-D reconstruction of the recoil tracks can be achieved, Monte 
Carlo studies indicate that a WIMP signal could be identified with high 
confidence from as few as 30 detected WIMP-nucleus scattering events.
\endabstract

To identify a WIMP signal, conventional searches
rely mainly on searching for an annual modulation in the rate of
nuclear recoils in the detector. This arises from the varying combined 
velocity vectors of the Earth's orbital
motion and the Galactic rotation~\cite{drukier}.  Because this is only a 5\% effect, and 
existing detectors generally have a large background from Compton 
electrons, many WIMP-nucleus scattering events are needed to identify a 
WIMP signal statistically.  Therefore
emphasis has mainly been placed on using very large mass detectors to increase the
event rate, and great 
long-term system stability is required.  In this paper, we present an 
alternative technique which, although it has a very low total target mass, 
has a strong signature for WIMPs and excellent background rejection 
capabilities.

\begin{figure}
\begin{center}
\leavevmode
\hbox{%
\epsfxsize=3.4in
\epsffile{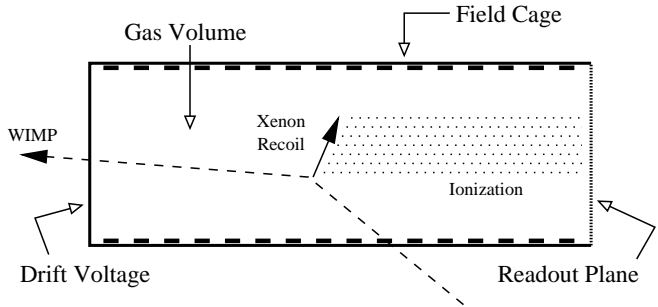}}
\end{center}
\caption{Schematic of TPC.}
\label{fig:detector}
\end{figure}

The DRIFT experiment (Directional Recoil Identification From
Tracks) is the successor to a project begun at UC San Diego~\cite{prl}.  The
detector is a Time Projection Chamber (TPC), consisting of a target
gas volume in a strong electric field (see figure~\ref{fig:detector}).  A nuclear recoil 
in the target volume will ionize the gas as it loses energy.  A low
pressure gas is used to extend the ranges of these
ionization tracks to a few mm for typical WIMP-induced recoil energies
($\sim$ 100 keV).  In the UCSD detector scheme,
the ionized electrons were then drifted by the electric field to an
optically imaged Parallel Plate Avalanche Chamber (PPAC) in the end cap.
The entire detector was placed in a 4.5 kG large bore superconducting 
magnet to suppress the diffusion of the electrons as they drifted through
the gas. This was necessary in order to maintain the resolution of the 
original track~\cite{diffusion}.  The need for the magnet made a large scale-up of the
project impractical and prohibitively expensive, so a new method of
suppressing the diffusion was needed.

A breakthrough was made when it was
realized that a slightly electronegative gas admixture (in this case,
CS$_2$) would reversably attach the ionized electrons~\cite{crane}.
The resulting negative ions then drift to the high field gain region where
they release the electrons and avalanche on the anodes. It was expected
that the diffusion of the ions would be much less than that of the
electrons; this has now been verified~\cite{martoff,tohru}.
There is the added advantage that the longitudinal diffusion is suppressed 
as well as the transverse diffusion, which is not the case for electrons
in a magnetic field.  The timing of the ions arriving at the readout plane 
therefore carries valuable information about the drift-direction extension
of the tracks.  Along with the transverse position, a three dimensional
reconstruction of the track becomes possible when only two (transverse)
dimensions were available using the magnetic field technique.  Our 
preliminary measurements~\cite{martoff,tohru} show that sub-millimeter resolution can be
maintained after drifting 1~m in 20~torr of a Xe-CS$_2$ (50:50) mixture, and 
it is hoped that further investigation will show that even longer drift
distances are possible.

After the ions are drifted 
to the end of the chamber, the timing and spatial distributions must be
determined in order to reconstruct the track.  In the UCSD work,
an optically imaged PPAC was used, but there were difficulties in 
achieving the necessary resolution.  Other readout schemes are being
investigated, including multiwire proportional chambers, pads,
microdots, microstrips, and GEMMs.  In the end we hope to achieve the 
sensitivity required to accurately determine the recoil direction  
(both its axis and the sense of the motion) and energy for tracks longer
than 2~mm.  This corresponds to an energy threshold of about 100~keV.
This resolution should allow virtually all Compton electrons to be 
rejected, since the electron range at given total
ionization is predicted to be many times longer than for a Xenon recoil.  
Measurements to confirm the electron rejection efficiency will soon be
undertaken at the University of Sheffield Neutron Beam Facility and at 
Occidental College.  Background 
due to alpha-emitting contaminants in the gain-producing electrodes
(wires, grids or strips) can also be rejected with good efficiency using
total ionization and tracking cuts~\cite{burwell}.
This superior background rejection
allows this detection scheme to be competitive with
conventional techniques with much larger target masses. In the gas detector
method, background events are rejected on an event-by-event basis, while
other methods rely on a statistical subtraction of background which
introduces larger uncertainties and thus lowers the overall sensitivity.

\begin{figure}[t]
\begin{center}
\leavevmode
\hbox{%
\epsfxsize=3.4in
\epsffile{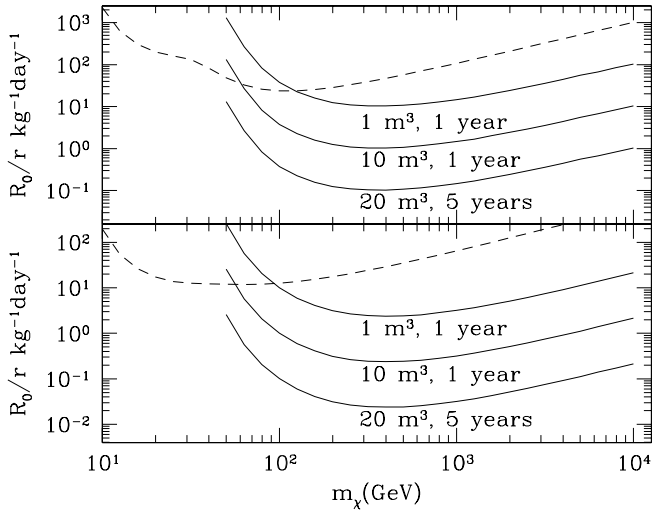}}
\end{center}
\caption{Upper limits (95\%~c.l.) on rate, normalized to Ge,
that may be achieved for coherent (top) and
spin dependent (bottom) WIMP-Xenon interactions
at different exposures, assuming zero background. Also shown with the
dashed lines are the current UKDMC NaI limits.}
\label{fig:limits}
\end{figure}

To determine the sensitivity with the above assumptions, a Monte Carlo simulation
was run in which WIMP-nucleus scattering events were generated
in a Xe-CS$_2$ (50:50) mixture and the ranges 
of the recoils were estimated using a software package entitled SRIM~\cite{trim}.
(SRIM calculations in low pressure gases are known to be off by as much as a factor of two,
and because this will dominate the uncertainty of these results, no diffusion or straggling
effects were included in the simulation. A range-energy calibration is planned for
early next year using the Sheffield neutron beam.)
The number of recoils above threshold was recorded and the detection
efficiency as a function of WIMP mass was determined.  A plot of the 
sensitivity of such a detector is shown in figure~\ref{fig:limits}.  The curves shown 
are the 
upper limits on rate that could be set if no nuclear recoils
were detected at various exposures.  Note that the limits would be
stronger than the current UKDMC NaI limits [5], even though the density
of the Xenon targets is only 0.07 kg/m$^3$.
This attests to the fact that high sensitivity can indeed be achieved with
a very low target mass if the backgrounds can be sufficiently
reduced.

\begin{figure}[t]
\begin{center}
\leavevmode
\hbox{%
\epsfxsize=3.4in
\epsffile{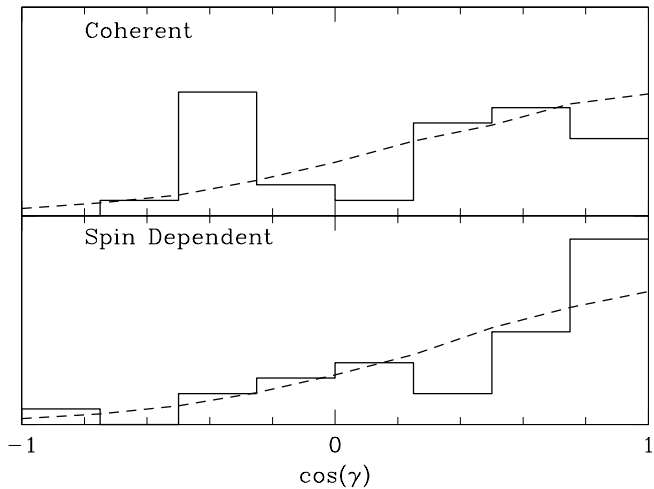}}
\end{center}
\caption{Histograms of recoil angles in Galactic frame (direction of sun
through the Galaxy is at $180^{\circ}$) for 30 WIMP-Xenon interactions.
Also shown in the dashed lines are
the distributions for large numbers of events.}
\label{fig:coshist}
\end{figure}

However, more important than the ability to set an upper limit
is the sensitivity for an actual positive WIMP signal 
detection.  This is where the ability to determine the recoil axis and
direction would give a significant advantage.  Spergel [6] has shown the
WIMP interaction rate as a function of recoil energy and angle (in the 
Galactic frame) to be
\begin{equation}
{dR \over dE_{\rm R}d\cos \gamma} \propto \exp
\left[{-(v_\odot \cos \gamma - v_{\rm min})^2 \over
v_{\rm halo}^2}\right]
\end{equation}
where $v_{\rm min}^2 = (m_{\rm N} + m_\chi)^2 E_{\rm R}/2m_{\rm N}m_\chi^2$
and $v_{\rm halo}^2 = 3v_\odot^2/2$.
This equation is used in the Monte Carlo
simulation to generate the recoil angle spectrum as well as to determine
the detection efficiency.  Figure~\ref{fig:coshist} shows typical recoil angle spectra 
for 30 random WIMP events with axis and direction determination.  Note
the resemblance to the true spectra even at these statistics.
Kolmogorov-Smirnov tests indicate that the signals shown are inconsistent 
with an isotropic background at the 99\% c.l.  Thus in the absence of
background, such a detector could discover a WIMP signal with
high confidence from only 30 detected events.  Even with zero background,
more than 12,000 events  would be needed to detect a 5\% annual
modulation at the same confidence level.

Plans for the project begin with imaging readout tests 
with the UCSD prototype (0.18 m$^3$) in negative ion drift mode, 
and operation of a 0.03~m$^3$ prototype at Occidental College.  At the
same time a 1~m$^3$ detector will be constructed.  By the year 2001 we
hope to have a 20~m$^3$ detector operating in Boulby mine, which after
5 years of running would allow a positive WIMP detection at 
rates above 0.3 events/kg/day for a 1000 GeV WIMP.

\bibliography{dark98}

\end{document}